\documentclass[12pt]{iopart}

\usepackage{graphicx}
\usepackage{xspace}
\usepackage[ruled,vlined]{algorithm2e} 
\SetAlFnt{\small}

\newcommand{\bra}[1]{
    \ensuremath{\left\langle #1 \right|}\xspace}

\newcommand{\ket}[1]{
    \ensuremath{\left|  #1 \right\rangle}\xspace}

\begin{document}

\title[Resource-aware System Architectural Model for Implementation of QBA]{Resource-aware System Architecture Model for Implementation of Quantum aided Byzantine Agreement on Quantum Repeater Networks}

\author{M. Amin Taherkhani$^1$, Keivan Navi$^1$, Rodney Van Meter$^2$}

\address{$^1$ Faculty of Computer Science and Engineering, Shahid Beheshti University, IR}
\address{$^2$ Faculty of Environment and Information Studies, Keio University, JP}
\ead{m\_taherkhani@sbu.ac.ir, navi@sbu.ac.ir, rdv@sfc.wide.ad.jp}

\vspace{10pt}
\begin{indented}
\item[] June 2017 
\end{indented}

\begin{abstract}
Quantum aided Byzantine agreement (QBA) is an important distributed quantum algorithm with unique features in comparison to classical deterministic and randomized algorithms, requiring only a constant expected number of rounds in addition to giving a higher level of security. In this paper, we analyze details of the high level multi-party algorithm, and propose elements of the design for the quantum architecture and circuits required at each node to run the algorithm on a quantum repeater network. Our optimization techniques have reduced the quantum circuit depth by 44\% and the number of qubits in each node by 20\% for a minimum five-node setup compared to the design based on the standard arithmetic circuits. These improvements lead to a quantum system architecture with 160 qubits per node, space-time product (an estimate of the required fidelity) $KQ \approx 1.3 \times 10^{5}$ per node and error threshold $1.1 \times 10^{-6}$ for the total nodes in the network. The evaluation of the designed architecture shows that to execute the algorithm once on the minimum setup, we need to successfully distribute a total of 648 Bell pairs across the network, spread evenly between all pairs of nodes. This framework can be considered a starting point for establishing a road-map for light-weight demonstration of a distributed quantum application on quantum repeater networks. 
\end{abstract}

\vspace{2pc}

\maketitle

%
%
%
%
%

\section{Introduction}
\label{sec:Introduction}
Besides the numerous attempts to implement a large-scale quantum machine for local and centralized applications such as efficient factoring~\cite{Shor94}, there are other vast areas of research on interesting and unique distributed quantum algorithms with no classical variants~\cite{Anne15, Ashly16,VanMeter14}. Perceived advantages with respect to security, time complexity or communication complexity are sufficient to motivate researchers to focus on design, implementation and optimization of the algorithms. Quantum Key Distribution (QKD) protocols~\cite{E91,BB84} and quantum leader election algorithms~\cite{Tani12} are a subset of these algorithms~\cite{VanMeter14}. However, although detailed, resource-aware analyses of monolithic algorithms are increasing, the literature for equivalent analysis of distributed algorithms remains sparse.   

A quantum approach for solving the classical problem of \textit{Byzantine agreement} is another important distributed quantum algorithm, proposed by Ben-Or and Hassidim~\cite{Ben05}. Their pure-theoretical distributed algorithm terminates in $O(1)$ expected  number of  rounds in the presence of a computationally unbounded, full information and adaptive adversary. There is no similar variant in classical deterministic or randomized solutions with these unique features at the same time. 

Distributed algorithms for solving the Byzantine agreement problem are crucial for designing fault-tolerant systems in many domains. These algorithms have broad applications in areas ranging from fault-tolerant real-time and online services~\cite{Cast02,Kotla07}, to secure and large-scale peer-to-peer services~\cite{Young13}.  

Although quantum sharing-based Byzantine agreement can be theoretically faster and more secure than the classical algorithms, a considerable gap has remained from the abstract layer to experimental layer. To fill this gap, we must address challenges ranging from finding a practical architecture for quantum processing elements in each independent node to consideration of a quantum-based infrastructure for communication known as a \textit{quantum repeater network (QRN)}, by using entanglement and teleportation\cite{VanMeter13}. Analysis of faults in end-nodes and network imperfections in comparison to the ideal model of the original algorithm (which assumes perfect computation in each node and a perfect point-to-point quantum communication link between each pair) increase the complexity of the problem.

We face some questions about the quantum architecture of the abstract algorithm in addition to computational and communication resources required for running this algorithm. In this work we have:
\begin{itemize}
	\item extracted the minimal architecture requirements for the quantum part of the QBA protocol, 
	\item proposed two optimization techniques for the architecture and circuit for a minimum setup of QBA with 5 nodes, and 
	\item estimated computation and communication costs for the minimum setup, including required fidelity. 
\end{itemize}
  
  
%


The remainder of the paper is organized as follows: In section~\ref{sec:Preliminary}, we review classical and quantum aided Byzantine agreement and quantum repeater networks. Also, the main criteria for analysis of the QBA algorithm on quantum repeater networks is presented in this section. A high level analysis of the QBA protocol is presented in section~\ref{sec:HighLevelAnalysis}. In section~\ref{sec:Design}, the overall architecture design is proposed. We provide required optimization techniques in section~\ref{sec:Implementation}. The assessment of results is shown in section~\ref{sec:Result}. Finally we conclude the paper in section~\ref{sec:Conclusion}. 


\section{Background}
\label{sec:Preliminary}
In this section, we review history and related background. We start with the standard classical Byzantine agreement and history of Byzantine-tolerant solutions. We continue with the first scalable quantum sharing-based Byzantine agreement proposed by Ben-or and Hassidim in~\cite{Ben05}. A general review of the network requirements for execution of the algorithm concludes this section.        	    

\subsection{Classical Byzantine agreement}
The history of the Byzantine agreement problem goes back to a proposal by Lamport et al. for defining a more sophisticated form of fault model with active and malicious behavior, now known as Byzantine faults~\cite{Lamp82}. Tolerating this form of the fault, which is stronger than fail-stop faults, requires more computation and communication resources in a distributed system. To analyze the behavior of this type of fault, Lamport et al. employed the colorful metaphor of a distributed system as a group of Byzantine generals arrayed around a city, trying to decide as a group whether to attack or retreat. It must be assumed that less than third of these generals are traitors (active fault as an internal adversary). The mission of the distributed generals will be successful if all the loyal generals agree on a unique command (attack or retreat) in the presence of inconsistent messages sent by traitors (Byzantine faults), otherwise the protocol fails. In this problem, all communication is done by messenger. From the communication point of view, Byzantine agreement is a form of reliable broadcast between multiple nodes in a network which supports point to point channels between nodes.

\subsubsection{Problem Definition and Conditions} 
A distributed system of $N$ processes $Process_{0},Process_{1}, ..., Process_{N-1}$ is required to agree on a binary decision based on their initial value. The protocol is executed in the presence of a malicious adversary who can access some of the faulty processes in the system. A Byzantine agreement protocol satisfy these three conditions: 
\begin{itemize}
	\item \textit{Agreement:} After running the protocol, all non-faulty processes $Process_{i}$ must agree on the same value. 
	\item \textit{Validity:} If all non-faulty processes start the protocol with the same initial value, then all non-faulty processes must agree upon that value.  
	\item \textit{Termination:} All non-faulty processes will certainly make a decision during execution of the protocol. 
\end{itemize}

\subsubsection{Models and Assumptions} 
Byzantine faults present a more complex type of fault in comparison to fail-stop (or crash) faults. Faulty components can be controlled by an adversary with malicious behavior and used to send arbitrary messages to mislead non-faulty processes. In addition, from the system level point of view, the faulty processes are active players inside the game. This also increases the complexity of the protocol compared to other protocols with trusted parties in the game such as key distribution protocols.

Solutions to the Byzantine agreement problem are always designed with respect to a chosen set of assumptions. The main categories with impact on the problems are the timing model of communication between processes,  the knowledge held by the adversary, the behavior and computational power of the adversary: 
              
\begin{itemize}
	\item \textit{Communication model:} One of the most important assumptions for solving the problem is related to timing of communication between processes in the distributed system. The communication model is divided in two categories: \textit{synchronous} and \textit{asynchronous}. In the synchronous model, all processes are allowed to send messages only in well-defined communication rounds. Separate communication and computation phases are considered in each round of the protocol. On the other side, in the asynchronous communication model, no common clock is required to order the messages in well-defined communication rounds.

	\item \textit{Behavior of the adversary:} The capability of the adversary to have static or dynamic behavior during the execution of the protocol is also another important criterion. The adversary is categorized into \textit{static} or \textit{adaptive}. In the static case the adversary chooses her faulty processes before running the protocol. An adaptive adversary can change her behavior change the set of the faulty processes during the execution of protocol. In the last model, the only limitation is the upper bound of the number of faulty processes.  

	\item \textit{Computational model:} Another important assumption in Byzantine agreement problem is on the computational power of adversary to misinform the non-faulty processes. This criterion classify the adversaries into \textit{computationally bounded} and \textit{computationally unbounded}.  

	\item \textit{Information model:} In the following of computational power of the adversary, the level of information she can access from distributed system also needs to be specified. In the first type, \textit{private channel}, the adversary cannot access all information generated and communicated during the execution of protocol. The stronger type is known as \textit{full information} which means the adversary knows all the internal state of the processes.
\end{itemize}

The set of assumptions changes the hardness of the problem. The upper bound on the number of faulty players is reported in~\cite{Lamp82}.   

\subsubsection{Approaches and Solutions}
Many researchers have studied Byzantine agreement and proposed solutions for classical computing. The classical algorithms are divided into \textit{deterministic} and \textit{randomized}. Weaknesses and limitations of deterministic Byzantine agreement algorithms have shifted the focus of research to classical randomized algorithms. The main advantages of the randomized algorithms are lower round complexity and stronger security.  
An important approach for solving Byzantine agreement uses the concept of a common coin in the protocol. This means a sufficiently random bit would be available for all the non-faulty processes in the distributed systems. By using this feature, the solution does not require a trusted third party (TTP), in contrast to the encryption-based approaches. For implementation of the common coin feature in the network, many random numbers are generated by the processes in each round. The random numbers need to be shared and reconstructed using a \textit{secret sharing} scheme. In this scheme a dealer shared her secret between all the players in the presence at most \textit{t} dishonest players. A general theoretical secret sharing procedure may not satisfy all the real conditions specially in our problem. The conditions are secret sharing in presence of the faulty dealer and unavailability of ideal broadcast channel. Therefore we need to focus on more special versions of secret sharing procedure as described below:       
          
\begin{itemize}
	\item \textit{Verifiable secret sharing:} In general, no assumption is made about the trustworthiness of the dealer. In Byzantine agreement, each process may play the role of dealer at some points in the procedure. Therefore, to ensure a recoverable secret is shared by the dealer, we need a verifiable version of the procedure. In verifiable secret sharing, the dealer passes a commitment procedure and the players agree on the recoverability of the secret. 
	
	\item \textit{graded verifiable secret sharing:} In general all VSS protocols require a reliable broadcast channel. This is the main motivation for solving a Byzantine agreement problem and cannot be directly employed in any solutions. Also the existence of a reliable broadcast channel is not a commonly accepted assumption in a network research. In these conditions, the gradecast version of VSS is executed. The idea of graded-VSS is presented by Feldman \& Micali in~\cite{Feldman97}. In this protocol, all the players execute a gradecast protocol to replace the broadcast channel with a weaker but still efficient version of agreement.       
\end{itemize}





\subsection{Quantum aided Byzantine agreement}
One of the first attempts to exploit quantum advantages in a weaker version of agreement (known as detectable broadcast) as a form of reliable broadcast is presented by Fitzi et. al. in~\cite{Fitzi01}. Although the work doesn't solve Lamport's original problem, the solution is suitable for a small-scaled distributed systems (with 3 nodes) in a detectable broadcast application instead of Byzantine agreement. 
\begin{table}[b!]
	\centering
		\begin{tabular} {| c || c || c || c |}
		\hline 
		   & Deterministic & Randomized & Quantum-aided \\ 
			 & ~\cite{Fischer81} & ~\cite{Joseph98} & ~\cite{Ben05}\\ \hline \hline
			Round  & 									&								&    \\
			Complexity   & $O(t)$ & $O(t), \Omega(\sqrt{\frac{N}{log N}})$ & $O(1)$ \\ 
			 & 									&								&  \\

		\hline
		\end{tabular}
	\caption{Round complexity of Byzantine agreement in presence of full-information adversaries with the upper bound of malicious players ($t=O(N)$)}
	\label{tab:RoundComplexity}
\end{table}

In this paper, the focus is on Ben-Or and Hassidim's algorithm. Ben-Or and Hassidim proposed a scalable solution for creating a quantum aided Byzantine agreement protocol by modification of Feldman \& Micali's classical probabilistic algorithm~\cite{Feldman97} to share and verify a known quantum state, instead of sharing and verifying classical random numbers~\cite{Ben05}. Ben-Or and Hassidim's algorithm is based on the following assumptions: 

\begin{itemize}
	\item There exists a full-duplex ideal quantum channel between each pair of players (end-nodes). Note that during execution of the algorithm we also need an ideal classical channel between each pair of the players. The classical channels are used to fill the need for the broadcast channel, therefore privacy is not necessary for those channels. 
	\item To tolerate an upper bound on the number of faulty players ($t<N/3$), the communication model needs to be synchronous. Each round consists of two separate phases: a communication phase and a computation phase. For the asynchronous case, they prove the effectiveness of their algorithm with an upper bound of $t<N/4$.  
	\item The adversary can be adaptive, full information and computationally unbounded.  
\end{itemize}

At the end of algorithm execution, the agreement between all non-faulty players, the validity of the output between them and termination of the algorithm (with the probability of 1) are guaranteed. The algorithm can be analyzed from two points of view: 
\begin{enumerate}
	\item Performance analysis under security assumptions: the algorithm requires a constant number of expected rounds in the presence of a full information adversary with upper bound of malicious players $t$ among $N$ players ($t < N/3$).   
Table~\ref{tab:RoundComplexity} shows the upper bound of the round complexity in classical deterministic~\cite{Fischer81}, classical randomized~\cite{Joseph98} and quantum-aided algorithms~\cite{Ben05}. 
	\item Security analysis under performance assumptions: as another view, the quantum algorithm is more secure than deterministic and randomized algorithm in the case of the lowest bound of round complexity. 
If we analyze the security of Byzantine agreement protocols for fixed-round algorithms, we get the result similar to Table~\ref{tab:SecurityQBA}. There is no fixed-round, deterministic algorithm with the strongest type of adversaries (adaptive, computationally unbounded and full information)~\cite{Fischer81}. For the case of the randomized algorithms, the best available algorithm suffers from the assumption of communication security between each pair of non-faulty nodes~\cite{Feldman97}. 
\end{enumerate}

\begin{table}[b!]
	\centering
		\begin{tabular} {| c || c || c || c |}
		\hline 
		   & Deterministic & Randomized & Quantum-aided \\ 
			&               &~\cite{Feldman97} &~\cite{Ben05} \\ \hline \hline
			Adaptive Adv.& 			 & Yes & Yes \\ 
			Unbounded Adv.& No Solution & Yes & Yes \\ 
			Full Info. Adv.&  & No & Yes \\ 
		\hline
		\end{tabular}
	\caption{Security of fixed-round Byzantine protocols against adaptive, computationally unbounded and full information adversaries}
	\label{tab:SecurityQBA}
\end{table}

We will describe the details of the algorithm in addition to its theoretical behavior in section~\ref{sec:HighLevelAnalysis}.  
	
\subsection{Quantum Infrastructure}
\begin{figure}[h!]
	\centering
		\includegraphics[width=0.75\textwidth]{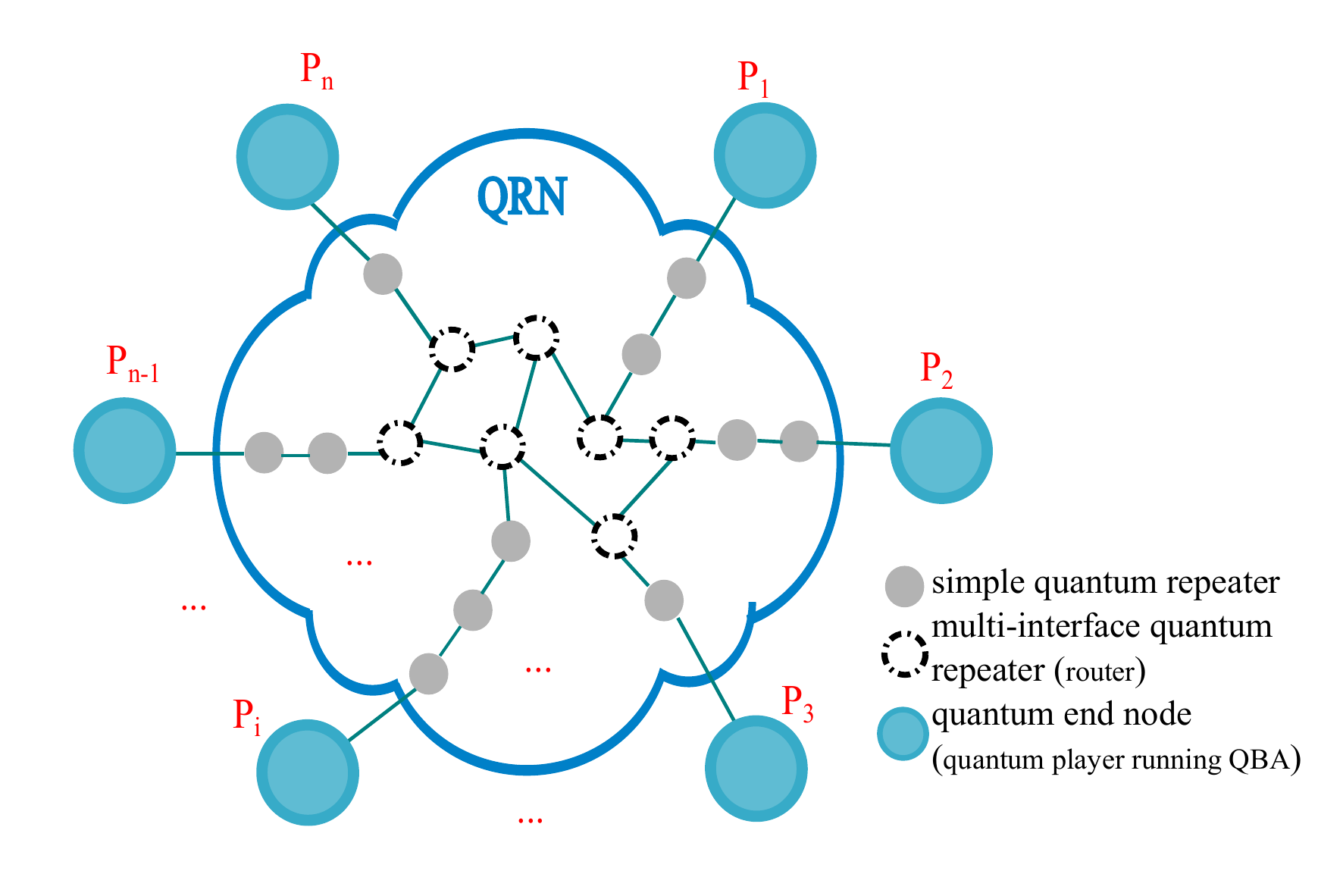}
	\caption{required elements in Quantum Repeater Networks (QRN) for running scalable quantum distributed applications}
	\label{fig:qrn-00}
\end{figure}

In general, execution of fundamentally distributed quantum algorithms requires a quantum-based solution for computation (at the end nodes) in addition to quantum-based communication known as a \textit{quantum repeater network (QRN)}~\cite{VanMeter14,VanMeter13,Kimble08,Hucul14,Ritter12} (Fig.~\ref{fig:qrn-00}). Quantum repeater networks provide an efficient infrastructure for distributed systems by using entanglement, teleportation and some forms of error detection and correction. In these networks, information could be represented by entangled states and a link between two quantum nodes creates entangled states supporting quantum teleportation. Purification and entanglement swapping repeaters~\cite{Dur99}, error correction-based repeaters~\cite{Jiang09,Fowler10} and quasi-asynchronous~\cite{Munro10} repeaters have been proposed for QRNs~\cite{Jiang16}. 

Due to the interdisciplinary nature of distributed fault tolerant protocols, distributed quantum algorithms and quantum networking, many concepts and criteria need to be clarified and defined before starting the detailed analysis. This section presents the basic definitions in these domains.

\subsubsection{Distributed Fault Tolerant Computing} 
In a distributed system, fault tolerant computation is modeled as a multi-party function evaluation in an unreliable environment. The unreliability comes from the error models in the presence of faults or errors. As we mentioned before, based on the error models the fault type can be classified as fail-stop or Byzantine.  

\subsubsection{Quantum Architecture and Circuit Design} 
In a general distributed quantum computation system, each node can execute quantum operations to complete a quantum protocol. We consider standard quantum circuit models. Also we need to consider the overall quantum architecture and the required circuits for each section of the protocol. Due to quantum resource limitations we have to remain aware of the need to design efficient circuits and architecture by evaluation of the following criteria in circuit models:     
\begin{itemize}
	\item Circuit depth ($K$): This parameter defines the maximum related operations required to complete the execution of a circuit treating one- and two-qubit gates as running in unit time. 
		
	\item Circuit qubits ($Q$): This parameter indicate the number of input/output and ancilla qubits required for preparation and quantum operations within the circuit.
	\item $KQ$: This product of $K$ and $Q$ defines an upper bound of the total number of quantum operations in a quantum circuit. 
	\item ($G_{total}$): The total number of quantum gates required for execution of the quantum circuit. We estimate this parameter by $G_{total} \approx KQ$.

\end{itemize}
 
\subsubsection{Quantum Networks Criteria} 
In addition to the local quantum resources in each node, we have to consider the communication resources and requirements. In general Bell pairs are the main resources in QRNs, with the following criteria:        
\begin{itemize}
	\item Bell pair Fidelity: This measure shows the level of perfection in Bell state during the quantum communication. We use  $F=\bra{\psi}\rho\ket{\psi}$, where $\ket{\psi}$ is a Bell pair and $\rho$ is the density matrix of the states created by QRN, so that infidelity $1-F$ is the error probability ~\cite{Jozsa94}. 
	
	\item Number of Bell pairs: During the communication rounds of a quantum protocol, each party may send some quantum shares by executing teleportation using the Bell pairs. We need to be aware of the number of Bell pairs consumed during the execution of the protocol. 
\end{itemize}

Despite many ongoing demonstrations of quantum key distribution (QKD) without the benefits of entangled repeater networks, the research on architecture analysis and design of other promising distributed quantum applications is rare and narrow. To the best of our knowledge, there is no detailed analysis and design for the quantum-aided Byzantine agreement algorithm. Therefore, 
there is a need for analysis of the minimum requirements in quantum repeater networks for complete execution of quantum Byzantine agreement. The modeling must be extended beyond that described by Ben-Or and Hassidim~\cite{Ben05}. In particular, the algorithm as proposed has been analyzed assuming only pure states, and without reference to the demands made on the repeater network. 

For the remaining sections, the focus will be on finding appropriate answers for the following questions: 
\begin{itemize}
	\item What are the required quantum resources for QBA protocol? 
	\item How resilient is it to network and gate error? 
	\item Is it practical and attractive in the real world?    
	\item Can we use QBA as an early demonstration application of quantum repeater networks? 
\end{itemize}

For exploration of an efficient solution, we need to analyze the relevant part of graded verifiable quantum secret sharing (VQSS)~\cite{Feldman97,Cre02} and the QBA protocol in detail, and determine the overall architecture of the quantum processing elements as will be presented in the next section. 

\section{Detailed Analysis}
\label{sec:HighLevelAnalysis}
In the previous section, we introduced the quantum aided Byzantine agreement protocol and enumerated its features in qualitative terms. In this section we begin the quantitative analysis of the algorithm. 

\subsection{Overall QBA Protocol}
\label{sec:QBA2QOCC}
\paragraph{} As shown in Algorithm~\ref{QBAAlg}, to run the quantum-aided Byzantine agreement (QBA), all the end-nodes run the overall agreement protocol with 3 sub-protocols. The overall protocol is based on the original randomized algorithm~\cite{Feldman97} and has a constant expected number of rounds independent of $N$. The sequential sub-protocols of quantum-aided Byzantine agreement are $P_{r}$, $P_{0}$ and $P_{1}$. At the beginning, all nodes $i$ are supposed to start the protocol with a classical input value $b_{i}$.

\paragraph{}
\label{sec:QOCCPR}
All nodes concurrently and independently execute the first sub-protocol ($P_{r}$). This sub-protocol advances toward an agreement for non-faulty and uncertain nodes. At the first step node $i$ distributes his classical input value $b_{i}$ to all other nodes in the network. In this step, each node sends in total $N-1$ classical bits and may receives $N-1$ classical values from other peers. Each node separately counts the received input values. If the sum of incoming bits was more than $2N/3$, she selects $b_{i}=1$ as her input value for the next sub-protocol. In the case the summation of incoming bit is less than $N/3$, the node selects  $b_{i} = 0$. In the uncertain condition ($>N/3$ and $<2N/3$), the nodes must run an oblivious common coin (OCC) procedure. In the algorithm the classical OCC is replaced with quantum aided oblivious common coin (QOCC). This is the only quantum part among all of these sub-protocols. As we discuss later, QOCC is a modification of the original oblivious common coin procedure for independent and random coin flipping.   

\paragraph{}
After re-evaluation of $b_{i}$, each node runs the other two sub-protocols $P_{0}$ and $P_{1}$ sequentially after $P_{r}$. They are pure classical protocols for biasing the outcome of the coin flipping procedure into zero or one respectively. It has been theoretically proven that if the non-faulty nodes are in agreement on 1 (or 0) at the start of $P_{1}$ (or $P_{0}$), all of them will successfully terminate the protocol with the same output~\cite{Feldman97}. Each node individually terminates the protocol and the other alive nodes retain the last values reported by nodes that have terminated for updating and re-evaluating their parameter during the execution of the sub-protocols.     


\enlargethispage{1\baselineskip}

\begin{algorithm}

\DontPrintSemicolon
\SetAlgoLined

\SetKwInOut{Input}{input}\SetKwInOut{Output}{output}
\Input{Each processes begins with an input bit $b_{i}$.}
\Output{Classical bit $d$ with a fairness $\rho$ all other non-faulty $Process_{i}$  successfully return the same value ($b_{i}=d$)}
\nl ($\forall$ processes $(i)$) $\forall j$: $b_{j} \leftarrow null$, 
     $TerminateNextRound \leftarrow FALSE$ \;  

\While{(1)} { 
	\nl SUB-PROTOCOL $P_{r}(b_{i})$  \\
	 
		$O(N)$ classical messages to distribute $b_{i}$ \; 
		\If{(!TerminateNextRound)} {
			Flip distributed quantum coin, $r_{i} \longleftarrow QOCC(N)$ \;
			$x \longleftarrow $tally(all received $b_{j}$) \; 
			
			\lIf{($x < N/3$)} {
				$b_{i} \longleftarrow 0$ \;
			} 
			\lElseIf{($x > 2N/3$)} {
				$b_{i} \longleftarrow 1$ \;
			} 
			\lElse{$b_{i} \longleftarrow r_{i}$
			} 
		} \Else {
			return $d \longleftarrow b_{i}$ \;
		}

	\nl SUB-PROTOCOL $P_{0}(b_{i})$  \emph{//Classical Protocol} \\
		$O(N)$ classical messages to distribute $b_{i}$ \; 
		\If{(!TerminateNextRound)} {
			$x \longleftarrow $tally(all received $b_{j}$ or most recent $b_{j}$) \;
			\lIf{($x < N/3$)} {
				send $b_{i} \longleftarrow 0$, TerminateNextRound $\longleftarrow$ TRUE\;
			} 
			\lElseIf{($x > 2N/3$)} {
				$b_{i} \longleftarrow 1$ \;
			} 
			\lElse{$b_{i} \longleftarrow 0$ \;
			}
		} \Else{
			return $d \longleftarrow b_{i}$ \;
		}

	\nl SUB-PROTOCOL $P_{1}(b_{i})$ \emph{//Classical Protocol}  \\
			$O(N)$ classical messages to distribute $b_{i}$ \; 
		\If{(!TerminateNextRound} { 
			$x \longleftarrow $tally(all received $b_{j}$ or most recent $b_{j}$) \;
			\lIf{($x < N/3$)} {
				$b_{i} \longleftarrow 0$ \;
			} 
			\lElseIf{($x > 2N/3$)} {
				send $b_{i} \longleftarrow 1$, TerminateNextRound $\longleftarrow$ TRUE \;
			} 
			\lElse{$b_{i} \longleftarrow 1$\;
			}		
		
		} \Else{
			return $d \longleftarrow b_{i}$ \;
		}
	}
\caption{Quantum aided Byzantine agreement protocol \label{QBAAlg}~\cite{Ben05}~\cite{Feldman97}}
\end{algorithm}


\subsection{QOCC} 
\label{sec:QOCCSec}
The goal of running the quantum version of OCC (QOCC), similar to its classical predecessor (OCC), is to generate the common random bit among $N$ nodes to reach agreement in the condition of uncertainty. In the classical version, this goal is achieved by sharing $N^2$ random numbers between nodes (each node $i$ acts as a dealer to share her random value for each node $j$, $1 \leq j \leq N$)  using graded verifiable secret sharing. Unlike classical OCC, QOCC doesn't require privacy for modeling classical channels. Instead, the protocol takes advantage of quantum channels to share a known quantum state. The main quantum parts of QOCC are shown in Algorithm~\ref{QOCC}. The procedure is based on the modifications suggested in~\cite{Ben05}. As shown in the algorithm, we focus on the quantum part of QOCC and all of the remaining classical computations for identification of faulty players remain unchanged based on~\cite{Feldman97}.

\begin{algorithm}
\DontPrintSemicolon

\KwData{NONE}
\KwResult{Each $Process_{i}$ acquires the classical outcome of a common quantum coin ($r_{i}$)} 
\nl (\textit{All processes $(i)$}) Select a dealer $D$ \; 
\nl (\textit{All processes $(i)$}) $FP_i \leftarrow null$ \emph{//List of faulty processes} \;
\nl $D$ prepares a state $\ket{\phi} \longleftarrow \frac{1}{\sqrt{N}}\sum^{N-1}_{i=0} \ket{i}$\;
\nl (\textit{All processes $(i)$}) ($FP_{i}, $$\ket{S_{0,0}^{(i)}}) \longleftarrow GradedQSV(D,N, \ket{\phi})$ \; 

\nl (\textit{All processes $(i)$}) $Value_{j,i} \longleftarrow \mathcal{M}(\ket{S_{0,0}^{(j)}})$ \;

\nl (\textit{All processes $(i)$}) Gradecast($Value_{j,i}$), Update($FP_{i}$) \;

\nl (\textit{All processes $(i)$}) ($\forall$ j) \If {($j \in FP_{i}$)} {$SUM_{ij} = BAD$ \;} 
\nl \Else {$SUM_{ij} = \sum_{k \notin FP_{i}} Value_{k,j}$ $mod$ $N$} \; 									
																								
\nl	\If{(\emph{there exists any} $j$ \emph{ such that } $SUM_{ij} = 0$)} {
			return  $ r_{i} \longleftarrow 0$ \;} 
\nl		\Else{return $r_{i} \longleftarrow 1$ }

\caption{Quantum Oblivious Common Coin (QOCC) \label{QOCC}}
\end{algorithm}

As shown in Algorithm~\ref{QOCC}, at the first step a node is selected to act as a dealer (D) for sharing the following quantum state~\cite{Ben05}: 

\begin{equation} \label{eq:1}
	\ket{\phi}=\frac{1}{\sqrt{N}}\sum^{N-1}_{i=0}\ket{i}
\end{equation}

In the above equation $N$ is the number of end-nodes in the distributed system and $\ket{\phi}$ is a known quantum state for producing sufficiently random numbers among the nodes. To avoid cheating from a faulty dealer, a sub-section of the quantum version of verifiable secret sharing (VQSS) has been employed.  

Independent of the requirements in QBA, verifiable quantum secret sharing (VQSS) of any unknown state of ($\ket{\psi}$) is a 3-phase protocol with sharing, verification and reconstruction sub-protocols~\cite{Cre02}. 
To generate a common coin between the distributed nodes, the first two phases of the protocol can be effectively employed with the difference of replacing gradecast instead of broadcast~\cite{Ben05}. We call this part of the graded VQSS graded quantum share and verify (GradedQSV). As we discuss in the next section, for each process the output of the GradedQSV will be an $N$-qupit register. For each process $i$, a qupit $\ket{{{Q_{P}}^{(i)}}_{(0,0,j)}}$ (see Eq.~\ref{eq:6}) is a random value held by process $j$ representing what she assumes to be the value that process $i$ has chosen for her. Measurement of each qupit will present a random value. Note that during the execution of GradedQSV, the faulty processes will be caught and the random values of non-faulty processes can not be affected by the faulty processes. 

The details of GradedQSV are discussed in the next section. 

\subsection{GradedQSV}   
\label{sec:QSVSec}
The quantum sharing and verification procedure for QOCC is fundamentally based on the sharing phase of VQSS~\cite{Cre02}. The only difference between the sharing phase of the original VQSS and GradedQSV is the replacement of the ideal broadcast channel in~\cite{Cre02} with the gradecast procedure in~\cite{Feldman97}. GradedQSV, as its name implies, is divided into two subroutines, graded sharing and verification:





\subsubsection{Sharing Phase}
\label{sec:SharingPhase}
\begin{figure}[h!]
	\centering
		\includegraphics[width=0.95\textwidth]{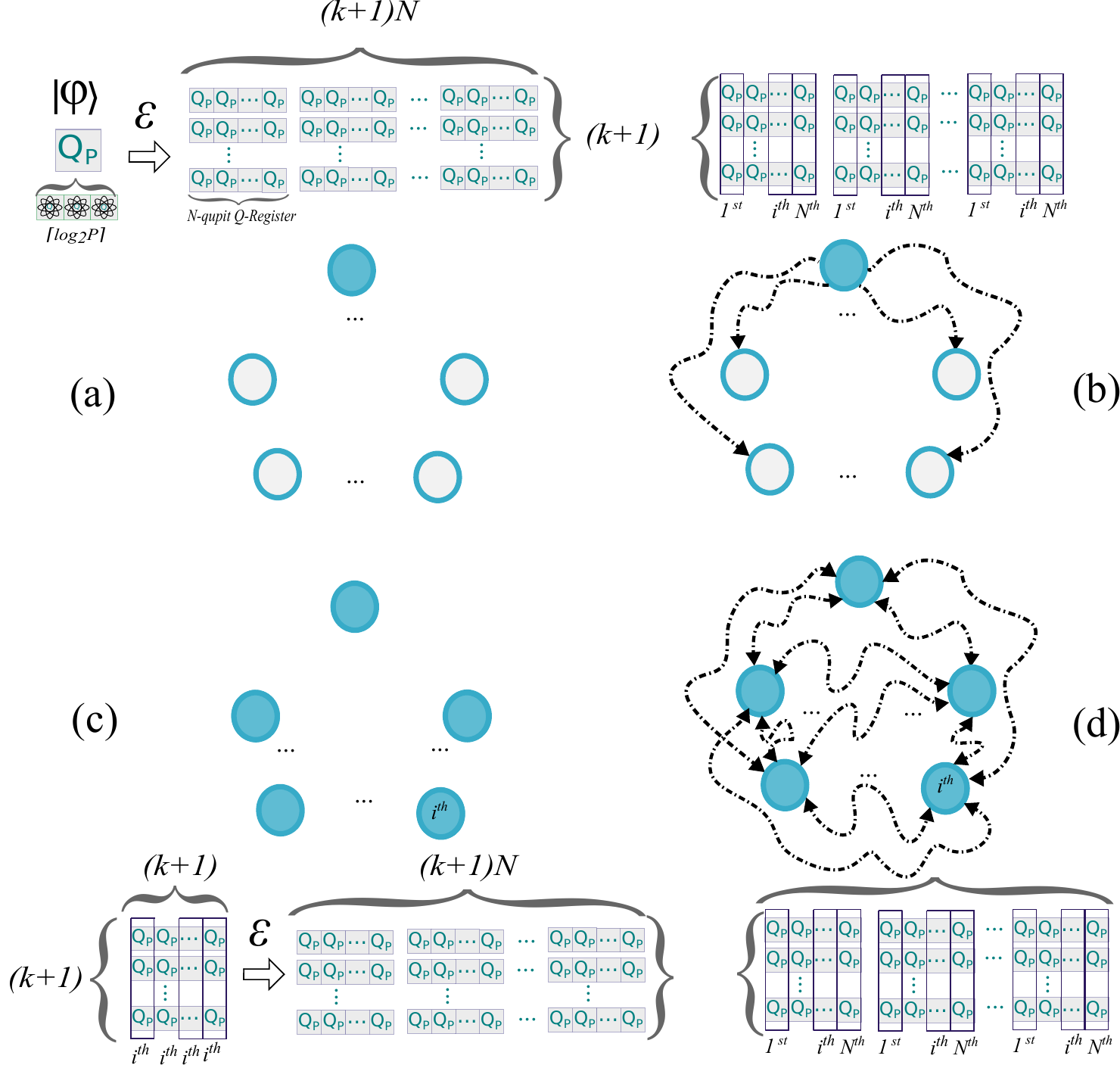}
	\caption{quantum state preparation and transmission during the sharing phase of gradedQSV for $N$ nodes and security parameter $k$. Each dot is a node, and each arrow represents transmission of a group of qupits by active nodes (filled colored nodes) via teleportation. The topmost node is the
dealer. (a) Dealer prepares and encodes initial quantum state using $N(k+1)^2$ qupits ($(k+1)^2$ N-qupit registers). (b) Dealer sends $i^{th}$ component of each N-qupit register to player $i$. (c) Each player ($i$) encodes each received qupit to a N-qupit register. (d) Each player ($i$) sends $j^{th}$ component of her encoded share to player $j$.      
}
	\label{fig:VQSS-Phases-01}
\end{figure}

The main quantum state preparation and communication for GradedQSV during the sharing phase is shown in Fig.~\ref{fig:VQSS-Phases-01}. 
The procedure as defined calls for P-level quantum variables (qupits) with the minimum prime number $P$ which is larger than number of nodes $N$ ($N < P$). In our implementation, we encode them in $\lceil \log _{2}{P}\rceil$ qubits, because Bell pairs support only distribution of qubits, and error correction circuits for addition, and physical systems are all best developed for two-level systems. 

The sharing phase includes some sub-phases with an agreed upon security parameter $k$. This parameter is a built-in engineering parameter for designing the VQSS-based protocols~\cite{Cre02}. The parameter improves the probability of capturing a dishonest dealer with probability on the order of $1-2^{-\Omega(k)}$.  
 
In the first level of the sharing phase, as shown in Fig.~\ref{fig:VQSS-Phases-01}-a, the dealer ($D$) prepares $(k+1)^2$ $N-$qupit quantum registers:   

\begin{equation} \label{eq:2}
\ket{S_{D_{(m,n)}}} = \bigotimes_{i=1}^{N}\ket{{{Q_P}^{(i)}}_{(m,n)}}  (0 \leq m,n \leq k)  
\end{equation}

In the above equation, $\ket{S_{D_{(i,j)}}}$ is the ${(ik+j)}$-th prepared system of the dealer ($(ik+j)$-th $N$-qupit register) and ${Q_{P}}^{(i)}_{(m,n)}$ is the $i$-th qupit of the system $\ket{S_{D_{(m,n)}}}$. 
  
The first $N$-qupit system of the dealer ($\ket{S_{D_{(0,0)}}}$) is prepared for encoding the original state $\ket{\phi}$ of the equation (\ref{eq:1}): 

\begin{equation} \label{eq:3}
\ket{S_{D_{(0,0)}}} = \scalebox{1.8}{$\varepsilon$}\ket{\phi} = \frac{1}{\sqrt{N}}\sum^{N-1}_{i=0} \scalebox{1.8}{$\varepsilon$}\ket{i}
\end{equation}
In the above equation, \scalebox{1.8}{$\varepsilon$}\ket{\phi} is the encoded state of the original state by using a quantum error correction code ($C$) such as the quantum Reed-Solomon code~\cite{Ahar97}. 



After that, the dealer assigns the following state to the next $k$ registers:  
\begin{equation} \label{eq:4}
\forall n, 0 <\forall n \leq k : \ket{S_{D_{(0,n)}}} = \sum_{a \in Z_{P}} \scalebox{1.8}{$\varepsilon$}\ket{a} = \scalebox{1.8}{$\varepsilon$}\ket{0} + \scalebox{1.8}{$\varepsilon$}\ket{1} + ... + \scalebox{1.8}{$\varepsilon$}\ket{P-1}  
\end{equation}

For all of the remaining $k(k+1)$ systems, the dealer initializes the registers with the state $\ket{\bar{0}}$: 
\begin{equation} \label{eq:5}
(\forall m, 1 < m \leq k),(\forall n, 0 \leq n \leq k) : \ket{S_{D_{(m,n)}}} = \ket{\bar{0}} 
\end{equation}

After state preparation, the dealer starts to send the ${Q_{P}}$ components of $\ket{S_{D_{(m,n)}}}$ systems qubits among the nodes. As shown in Fig.~\ref{fig:VQSS-Phases-01}-(b), for each node $i$, ${Q_{P}}^{(i)}_{(m,n)}$ of the system $\ket{S_{D_{(m,n)}}}$ is transmitted to the player by the dealer.    

The sharing phase continues by encoding and sharing the received qubits between all nodes. After receiving qupits ${Q_{P}^{(i)}}_{(m,n)}$, each player $i$ encode the qupits to $(k+1)^2$ $N$-qupit systems $\ket{S_{i_{(m,n)}}}$ (Fig.~\ref{fig:VQSS-Phases-01}-(c)):       

\begin{equation} \label{eq:6}
 (0 \leq m,n \leq k) : \ket{S_{i_{(m,n)}}} = \scalebox{1.8}{$\varepsilon$}\ket{{{Q_{P}}^{(i)}}_{(m,n)}} = \bigotimes_{j=1}^{N}\ket{{{Q_{P}}^{(i)}}_{(m,n,j)}}   
\end{equation}
 In the above equation, $\ket{S_{i_{(m,n)}}}$ and ${Q_{P}^{(i)}}_{(m,n)}$ are respectively the ${(mk+n)}$-th quantum register of the node $i$ and the $i$-th qupit of the system $\ket{S_{i_{(m,n)}}}$, and $\scalebox{1.8}{$\varepsilon$}$ is the same encoding scheme used in the previous steps.

By applying the sharing phase, a two level tree has been created and distributed among the nodes. This tree can be effectively used for verification of the original state in the next phase.

\subsubsection{Verification Phase}
\label{sec:VerificationPhase}

\begin{figure}[h!]
	\centering
		\includegraphics[width=0.60\textwidth]{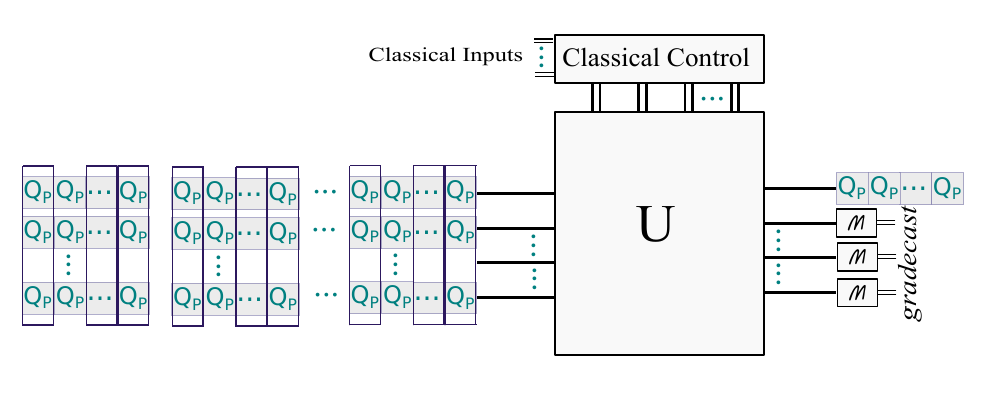}
	\caption{An overview of required inputs and output results in block-level verification diagram. A single quantum state is retained while the other are measured and their values shared via classical gradecast.}
	\label{fig:VerificationCircuit}
\end{figure}

As we show in the sharing phase, after the dealer encodes and shares the known state, the other nodes also apply another layer of encoding. In the next step, all of the non-faulty nodes act to increase their confidence about the correctness of the shared state. To do this, all of the nodes except the dealer independently execute the \emph{verification phase}. In this phase, each node applies some local operations on its share of the state, then measures some of their qubits and distributes the measurement results to all of the nodes, including the dealer, using a classical gradecast protocol. The general view for required assumptions about the inputs and the output results in quantum verification phase is shown in Fig.~\ref{fig:VerificationCircuit}. 

\subsection{Measurement}
\label{sec:MeasurementPhase}
   After execution of GradedQSV, the final measurement step is executed in all nodes.
	In this step, each node measures the remaining qupits it
holds. This will give all nodes sufficiently random numbers, which are
then shared via classical communication. This is the last quantum stage of
QOCC.

GradedQSV and the final measurement operations in QOCC are the only quantum 
portions of the complete QBA algorithm.
In the next section, we describe a more complete software architecture
for the quantum part of QBA, analyzed to establish hardware requirements in Sec.~\ref{sec:Result}.

\section{Architecture}
\label{sec:Design}
 
\subsection{Overall Design}
To establish the node architecture, we extracted the requirements for a minimum quantum setup for QBA. We require 5 nodes and need to set the prime number $P$ to be 7 for this setup. The security parameter ($k$) is selected to have the value two. These parameters lead to a minimum classical-quantum distributed system with five nodes which tolerates one malicious node inside the system. 

We target a resource-aware design for the proposed architecture. In this approach, the focus is to minimize the  quantum resources for implementation of QBA protocol. In view of the fact that a built-in quantum error correction is employed in QBA, we improved the design without using another layer of QEC in the architecture. Instead of using additional costly QEC, we look for an estimation of required gate error ($\epsilon_{g}$) to implement the protocol by considering the following equation:
\begin{equation}
	\epsilon_{g} << 1/G_{total} \approx 1/KQ 
\end{equation}

This means that for running QBA without using explicit QEC in the architecture, we need to demonstrate the protocol with gate error lower than $\epsilon_{g}$. 

The communication design of the system for execution of QBA protocol is based on the following assumptions: 
\begin{itemize}
	\item Both reliable classical and quantum channels exist between each pair of nodes. The classical channels are used in QBA to distribute the measured results. In contrast to the necessity of private classical channels in the classical approach, the no-cloning theorem of quantum states helps QBA to ignore the privacy assumption of the classical channels.  
	\item For quantum communication, the communication resource is Bell pairs. We focus on Bell states instead of W state or GHZ states in consideration of the capabilities of quantum repeater networks. 
	\item The Bell pairs may have fidelity $F < 1.0$.     
\end{itemize}

The above assumptions result in a complete graph topology. Although the protocol requires an expected constant number of rounds to be successfully completed, each round is computationally intensive, in addition to the high rate of quantum and classical communications. This requires long-lived connections between each pair of nodes. 


\subsection{Quantum Encoder} 
\begin{figure}[hbp!]
	\centering
		\includegraphics[width=0.75\textwidth]{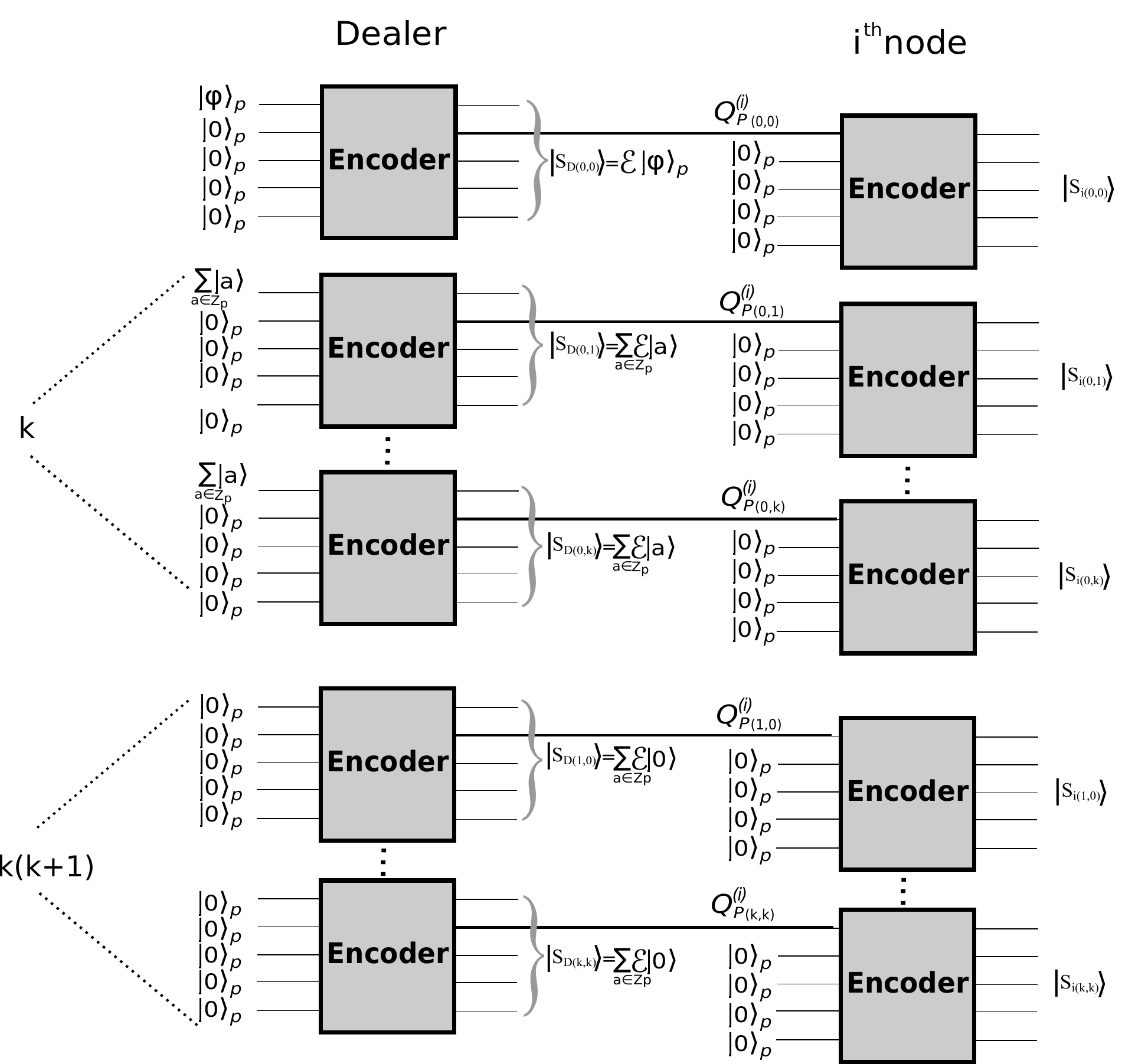}
	\caption{
	Block-level circuit diagram for the sharing phase of GradeQSV,
showing the Dealer and the circuits it executes on the left, and only
one of the $N$ players, $i$, on the right. On the left, the dealer
begins with the qupit $\ket{\phi}_P$ in the top line (\ref{eq:1} and line 2
of Algo.~\ref{QOCC}), $k$ superimposed numeric state qupits, and $k(k+1)$ zero
state qupits.  It executes $(k+1)^2$ Encoder circuits, then teleports
one qupit from the output of each Encoder to player $i$.
Each player creates additional ancillae and applies the Encoder to
\emph{each} qupit it received from the dealer.  Of the resulting
$N^2(k+1)^2$ qupits on the right hand side, each
player will keep $(k+1)^2$ and teleport the rest to the other $N-1$
nodes (Fig. 2-(d)).
}
	\label{fig:Encoder-Arch-01}
\end{figure}

Since the encoding and sharing scheme for the data qubits is generally similar for the dealer and the other  players, the general architecture is the same. Each player applies the encoding scheme as shown in Fig.~\ref{fig:Encoder-Arch-01}. As described in the previous section, for the sharing phase we need two-level encoding. The first one is applied by the dealer and the last one is applied by other players.     

The sharing phase of GradedQSV consists of a large number of calls to
the Encoder circuit, and a large number of teleportations to share the
qupits. For each player $i$, the Encoder circuit takes a qupit to be encoded and $N-1$
$\ket{0}_P$ ancillae, and creates the state $\ket{S_{D_{(m,n)}}}$. As described in
Sec.~\ref{sec:SharingPhase}, the dealer creates $(k+1)^2$ qupits, corresponding to Equations~\ref{eq:1}-~\ref{eq:6}, each of which are now run through the encoder.

\subsection{Quantum Verifier} 
Before the verification phase, each player has distributed the encoded qubits and in this phase the non-faulty nodes can verify the correctness of the original share (encoded by the dealer). In this phase, we need to extract the required quantum modules for unitary transform $U$ of Fig.~\ref{fig:VerificationCircuit}. In general, for every player $i$ unitary transform $U$ is composed of $N$ separated verifiers with the unitary transform $U^{(i,j)}$. Each $U^{i,j}$ is also decomposed to $(k+1)(k+2)$ transforms $U_{m,n}^{(i,j)}$ ($0 \leq m \leq k+1$, $0\leq n \leq k+1$). 

Here, $U_{m,n}^{(i,j)} = CX^{b_{m+1}}$. For all $m,n$ such that ($0 \leq m \leq k-1 , 0 \leq n \leq k$) we have:   
\begin{equation} \label{eq:verify1}
\ket{{{Q_{P}}^{(i)}}_{(n,0,j)}}\ket{{{Q_{P}}^{(i)}}_{(n,m+1,j)}} \\
\rightarrow CX^{b_{m+1}}\ket{{Q_{P}}^{(i)}_{(n,0,j)}}\ket{{Q_{P}}^{(i)}_{(n,m+1,j)}} \\
\end{equation}
    
In the above equation, $b_{l}$ are the classical values (for classical control) and $CX^{b_{l}}$ is modular multiplication and addition with the following described as below~\cite{Cre02}:
\begin{equation} \label{eq:verify2}
	CX^{b_{l}}\ket{V}\ket{W}=\ket{V}\ket{V+b_{l}W \bmod  P}
\end{equation}

For the case of $m = k$ and $\forall n$, $0 \leq n \leq k$, $U_{m,n}^{(i,j)}$ is determined as below:    
\begin{equation} \label{eq:verify3}
U_{m,n}^{(i,j)}: \ket{{{Q_{P}}^{(i)}}_{(n,0,j)}}
\rightarrow QFT(\ket{{Q_{P}}^{(i)}_{(n,0,j)}}) = \ket{{Q^{*}_{P}}^{(i)}_{(n,0,j)}}
\end{equation}

$U_{m,n}^{(i,j)}$ is defined as below for the case of $m = k+1$ and $\forall n$, $0 \leq n < k-1$:
\begin{equation} \label{eq:verify4}
U_{m,n}^{(i,j)}: \ket{{{Q^{*}_{P}}^{(i)}}_{(n,0,j)}}\ket{{{Q^{*}_{P}}^{(i)}}_{(n+1,0,j)}} 
\rightarrow CX^{b^{\prime}_{n+1}}\ket{{Q^{*}_{P}}^{(i)}_{(0,0,j)}}\ket{{Q^{*}_{P}}^{(i)}_{(n+1,0,j)}} 
\end{equation}

And finally unitary transform of $U_{k+1,k}^{(i,j)}$, ($m=k+1,n=k$) is applied to reverse the Fourier transform as below:

\begin{equation} \label{eq:verify5}
U_{k+1,k}^{(i,j)}: \ket{{Q^{*}_{P}}^{(i)}_{(0,0,j)}} \rightarrow QFT^{-1}(\ket{{Q^{*}_{P}}^{(i)}_{(0,0,j)}}) = \ket{{Q_{P}}^{(i)}_{(0,0,j)}} 
\end{equation}

\begin{figure*}
	\centering
		\includegraphics[width=0.90\textwidth]{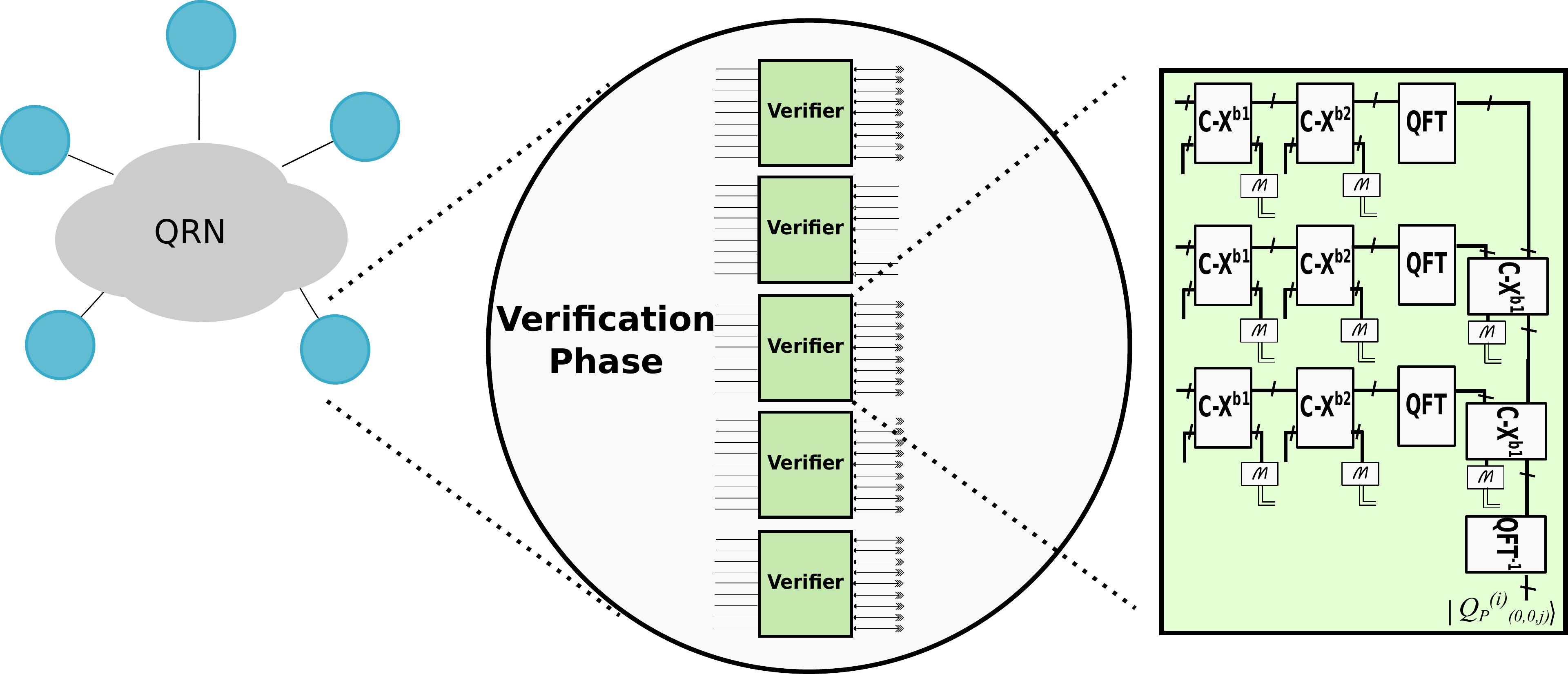}
    \caption{In the verification phase, the verifier circuit is
run on the collection of qupits received during the sharing phase.}	
	\label{fig:VerificationArch-02_1}
\end{figure*}

The required architecture for each node in a network with $N=5$ nodes is shown in Fig.~\ref{fig:VerificationArch-02_1}. Each player is required to have a quantum verifier module (each contains $U^{(i,j)}$) for every other node in the network. As shown in the figure, $CX^{b}$ is one of the main parts of the quantum verifier module. In the next section we will present an efficient multiplier mod 7 for reducing the cost of this module. The other main components of the quantum verifier consist of the quantum Fourier transform ($QFT$) and its inverse ($QFT^{-1}$). The required module for these operations is presented in Fig.~\ref{fig:VerificationArch-02_1}. Note that in the middle of verification phase, we required the measured outputs of $CX^b_{i}$ circuits. As shown in Fig~\ref{fig:VerificationCircuit}, the results of measurements are classically gradecast to verify the recoverability of the encoded original state $\ket{\phi}$. The remaining share  ${Q_{P}}^{(i)}_{(0,0,j)}$ are reserved for the final measurement described in the  next section.

\subsection{Measurement} 
After finalizing GradedQSV, and in the last stage of QOCC, the only internal operation is measurement of the remaining qubits. Since no other quantum gates are required, it may be considered to be the simplest part of the protocol. 
But maintaining the remaining qubits is considered to be an important challenge for this phase. The result of the measurement will be gradecast by the corresponding node.     


\section{Efficient Implementation} 
\label{sec:Implementation}
In this section we present some techniques for implementing and optimizing the quantum architecture level and quantum circuit level. For the first level, we apply resource sharing in a pipelined approach, and for the last one we propose an efficient low-cost modulo 7 classical-quantum multiplier.      

\subsection{Pipeline and resource sharing} 
\begin{figure}[b!]
	\centering
		\includegraphics[width=0.75\textwidth]{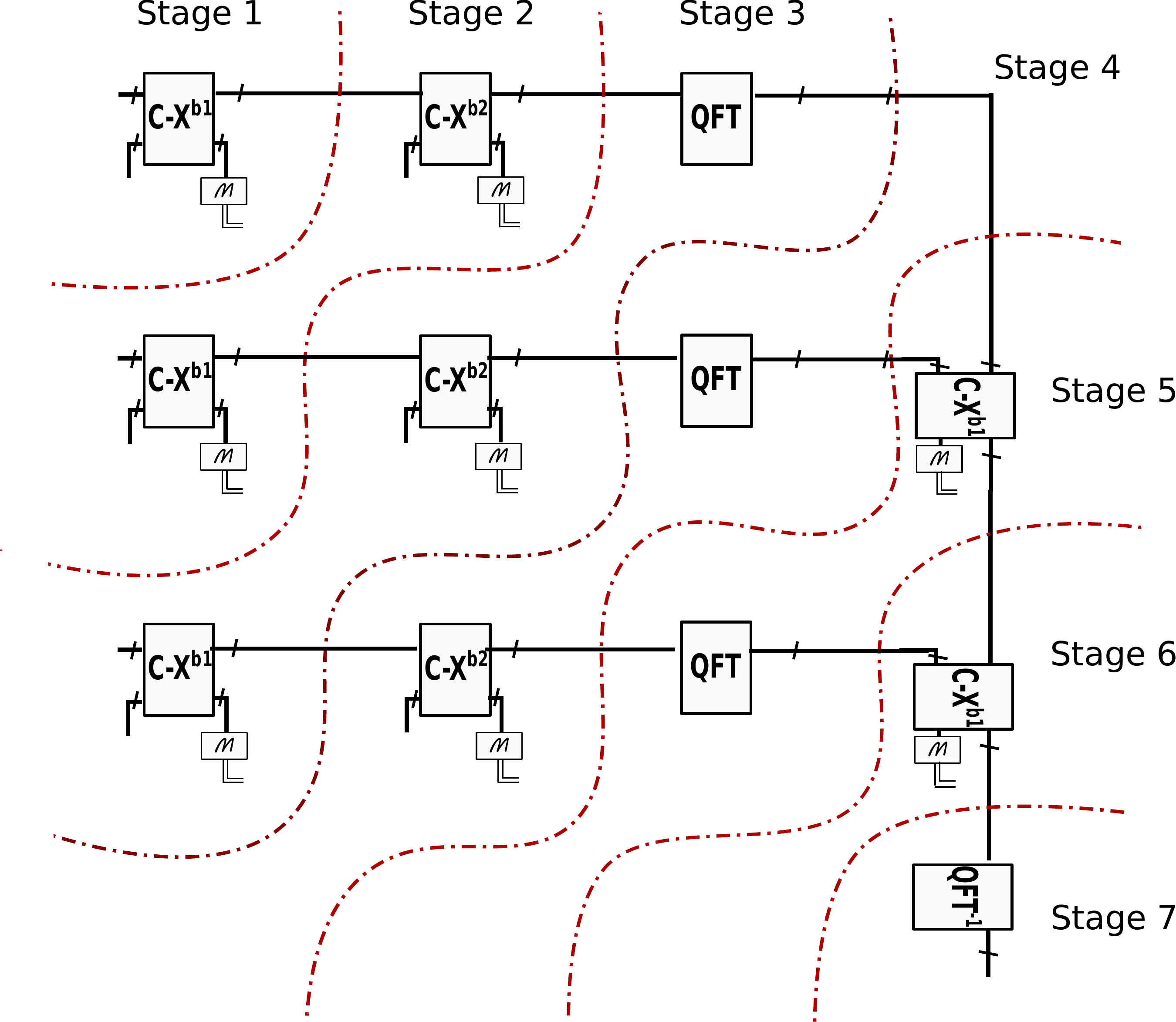}
		\caption{Parallel processing of quantum operation in verification phase for reduction of quantum cost and the number of required active qubits.}
	\label{fig:Parallel-01}
\end{figure}

To reduce the number of required quantum operations, reusing the quantum modules should be considered. In this technique, it is not required that all the modules be available at the same time. In the verification circuit, since multiplication and addition ($CX^{b}$) and $QFT$ ($QFT^{-1}$) are dependent on complete execution of the previous $CX^{b}$ and $QFT$ ($QFT^{-1}$) circuits, a resource sharing scheme has been designed to reduce the number of quantum modules and active number of qubits. As shown in Fig.~\ref{fig:Parallel-01}, we divide the execution time into 7 stages and in each stage a subsection of the required quantum operations is bound to the designed quantum modules. Although the overall execution time is increased by one stage in comparison to full circuit architecture execution time, the number of required quantum modules decreased dramatically. By applying this technique, in each verification module and in the worst case, we need to implement two $CX^{b}$ and one $QFT$ and/or $QFT^{-1}$ concurrently, instead of 8 active $CX^{b}$ modules, 4 active $QFT$ and $QFT^{-1}$ modules. 

In addition to reduction of the number of quantum modules, this optimization technique can reduce the required number of active qubits in the architecture.

\subsection{Multiplication Mod 7} 
As shown in the previous section, the most challenging computational part of the quantum modules is the circuit for computation of multiplication and addition modulo $P$ ($CX^{b}$). The concept for standard design of quantum circuit for computation of modular multiplication is based on modular addition. Motivated by designing efficient modular multiplication for the integer factorization application, many researchers have proposed low cost quantum circuits~\cite{VBE96,CDKM04,VanMeter05}. The standard approach for computation of $\ket{x} \ket{0} \rightarrow \ket{x} \ket{ax\: \bmod \: N}$ (for classical integers $N$ and $a$ and quantum variable $x$) is using computation of modular addition $\ket{x}\ket{y} \rightarrow \ket{x} \ket{x + y\: \bmod\: N}$. 

\begin{figure}
	\centering
		\includegraphics[width=0.75\textwidth]{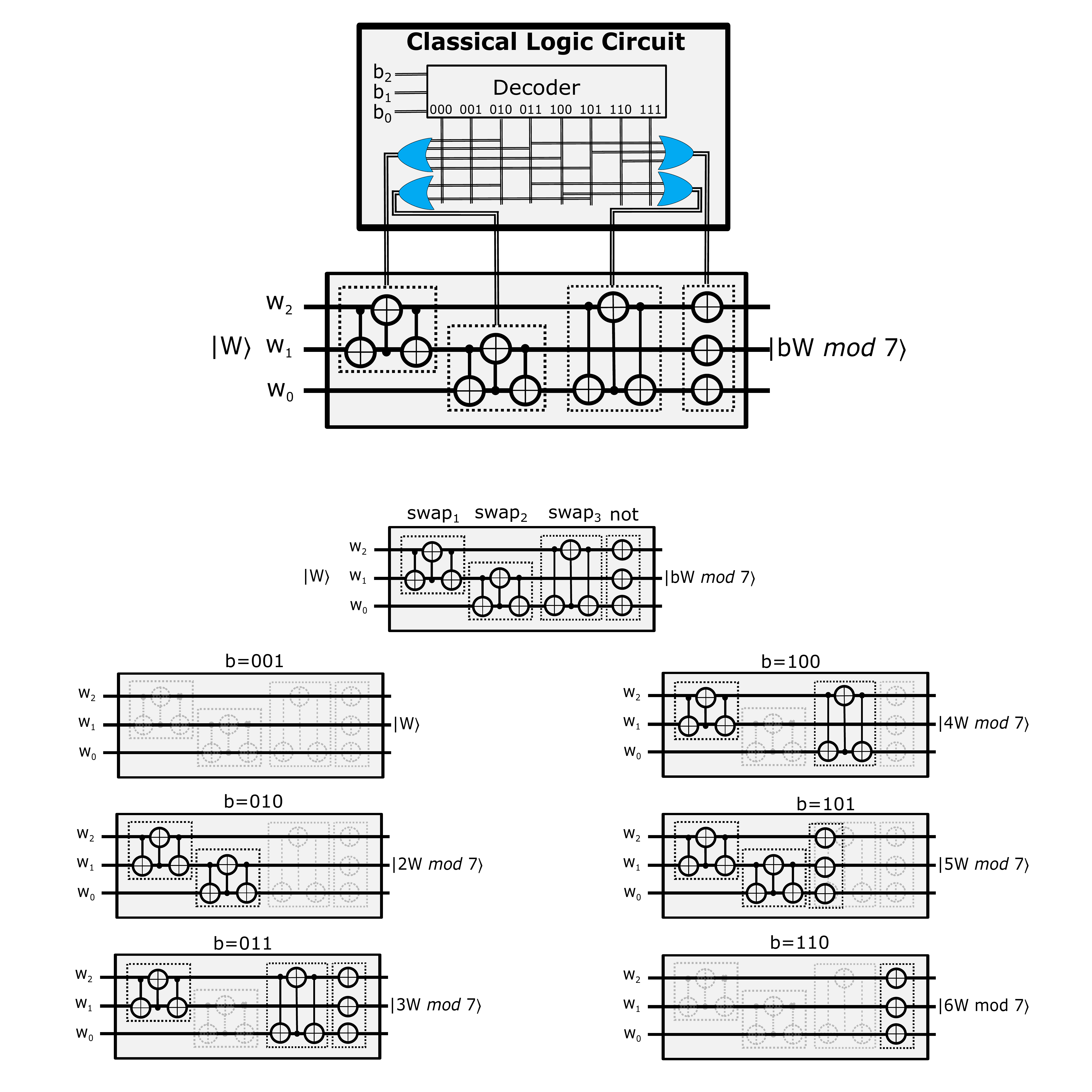}
	\caption{Multiplier modulo 7}
	\label{fig:ProposedMultiplier-03}
\end{figure}

Although this problem is similar to the state of the art modular multiplication (and modular addition), the concept of computation and the resulting circuit may be different from the state of the art computations and related circuits. The first difference is related to modulo computation. For integer factorization, the modulo $N$ is a parameter and can be changed for each new session. In contrast, the modulo $P$ depends on the upper bound of number of the players in QBA protocol, and can be considered a fixed number for a designed quantum distributed system. The next difference is related to the classical operand $a$ in integer factorization and classical random numbers $b_{i}$ in modular multiplication and addition circuit of the verification module. For the first one the operand can be implicitly considered in the circuit design, but for the last one, the random number must be present as a parameter for the modular multiplication. These differences cause a need for rethinking novel quantum circuits to be used in the QBA protocol. Thus we have proposed a multiplier modulo 7 for the minimum setup. We haven't followed the standard approach in integer factorization and attempted to compute the multiplication modulo 7 directly. The proposed modular multiplier is shown in Fig.~\ref{fig:ProposedMultiplier-03}. Note that $0 < b < 7$ in the protocol and the enable classical signals for 3 swap gates ($EN_{Swap_{i}}$) and not gates ($EN_{Not}$) are based on the following Boolean functions:
	\[EN_{Swap_{1}} = \bar{b_{2}}b_{1}\bar{b_{0}}+\bar{b_{2}}b_{1}b_{0}+b_{2}\bar{b_{1}}\bar{b_{0}}+b_{2}\bar{b_{1}}b_{0}
\]

\[EN_{Swap_{2}} = \bar{b_{2}}b_{1}\bar{b_{0}}+b_{2}\bar{b_{1}}b_{0}
\]

\[EN_{Swap_{3}} = \bar{b_{2}}b_{1}b_{0}+b_{2}\bar{b_{1}}\bar{b_{0}}
\]

	\[EN_{Not} = \bar{b_{2}}b_{1}\bar{b_{0}}+b_{2}\bar{b_{1}}\bar{b_{0}}+b_{2}b_{1}\bar{b_{0}}
\]

\section{Results for the minimum setup}
\label{sec:Result}
In this section the required quantum resources have been analyzed and enumerated for the minimum setup described in section~\ref{sec:Design}. For the analysis the following criteria have been considered: 
  
	\begin{itemize}
		\item Number of qubits per node
		\item Number of quantum operations (including total quantum cost and circuit depth)
		\item Number of Bell pairs consumed by the application	
	\end{itemize}

\subsection{End-node Quantum Cost}
\label{sec:EndNodeQuantumCost}

For analysis of quantum cost in the end-nodes, we consider the basic quantum circuits to be used in the architecture. The most complex quantum operations are in the verification phase. We used the basic quantum circuits for 3-qubit $QFT$ and $QFT^{-1}$. As we described in the previous section, the most challenging operation is related to modular multiplication and addition ($CX^{b}$). 
We consider five different designs for analysis of the quantum depth and total quantum cost in the verifier module: 
\begin{itemize}
	\item \textit{VBE96:} use the adder, modulo adder and modulo multiplier proposed in~\cite{VBE96}. 
	\item \textit{CDKM04:} replace the the basic adder of~\cite{VBE96} by the second adder proposed in~\cite{CDKM04}.
	
	\item \textit{VI05:} use quantum addition and modulo addition circuits of CDKM04. The only difference is replacement of the modulo multiplier in~\cite{VBE96}  by the multiplier proposed in~\cite{VanMeter05}. 
	\item \textit{Custom:} use the proposed multiplier modulo 7 instead of the previous standard design.
	\item \textit{Custom+Pipelined:} apply the proposed pipeline technique and using the circuit in the previous design.     
\end{itemize}

To evaluate the total cost of the required quantum operations, we estimate the required quantum gates based on the cost of CNOT as the basic gate. 

\begin{figure}[h!]
	\centering
		\includegraphics[width=1.00\textwidth]{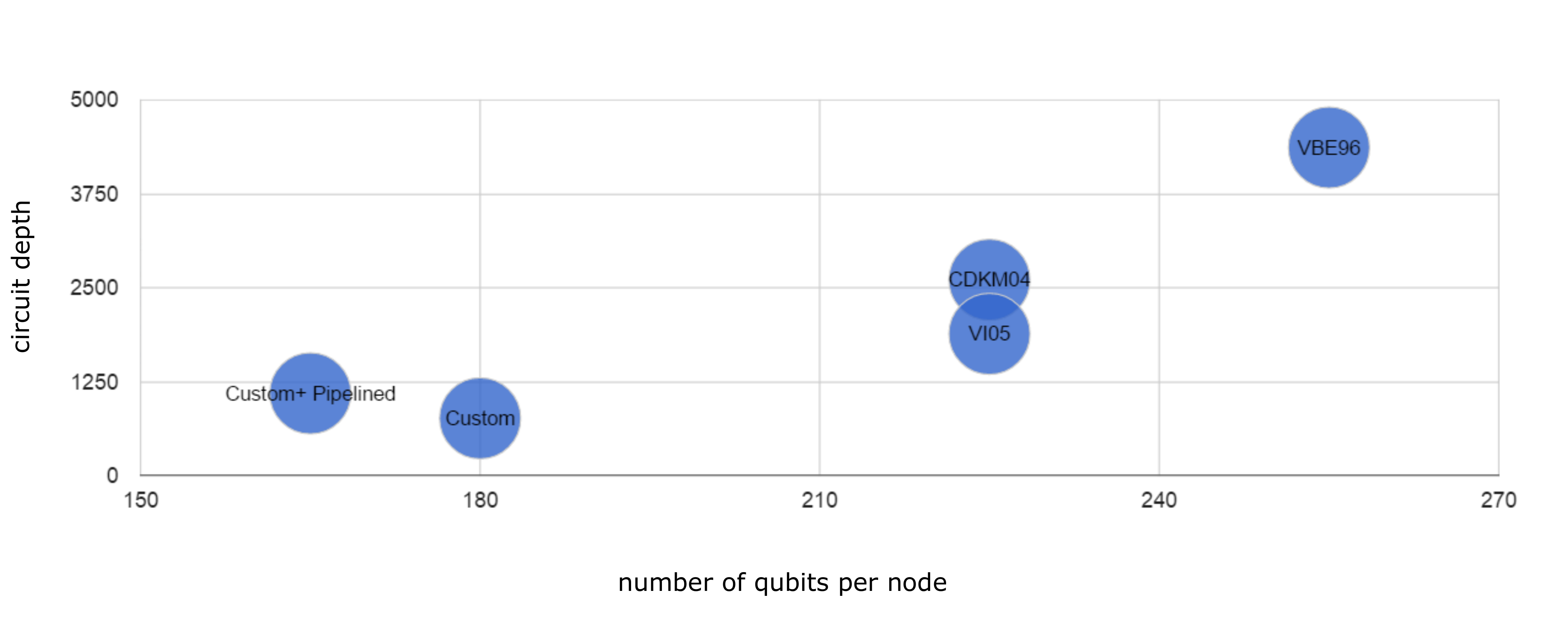}
	\caption{The number of required qubits per node versus the total required depth of the quantum circuit for the different designs. Points toward the lower left are better.}
	\label{fig:Result-Both-01}
\end{figure}

Due to reduction of the number of ancilla by using the proposed multiplier modulo 7, we gain at least 20\% improvement in this parameter. Fig~\ref{fig:Result-Both-01} plots the number of qubits and total depth of the quantum circuit.   


In this paper, except the built-in quantum error correction ($C$), we did not employ any other quantum error correction scheme for the architecture because the target was to present a lightweight and simple design. This architecture can be effectively optimized to be demonstrated in experiments. 

\begin{figure}[t!]
	\centering
		\includegraphics[width=1.00\textwidth]{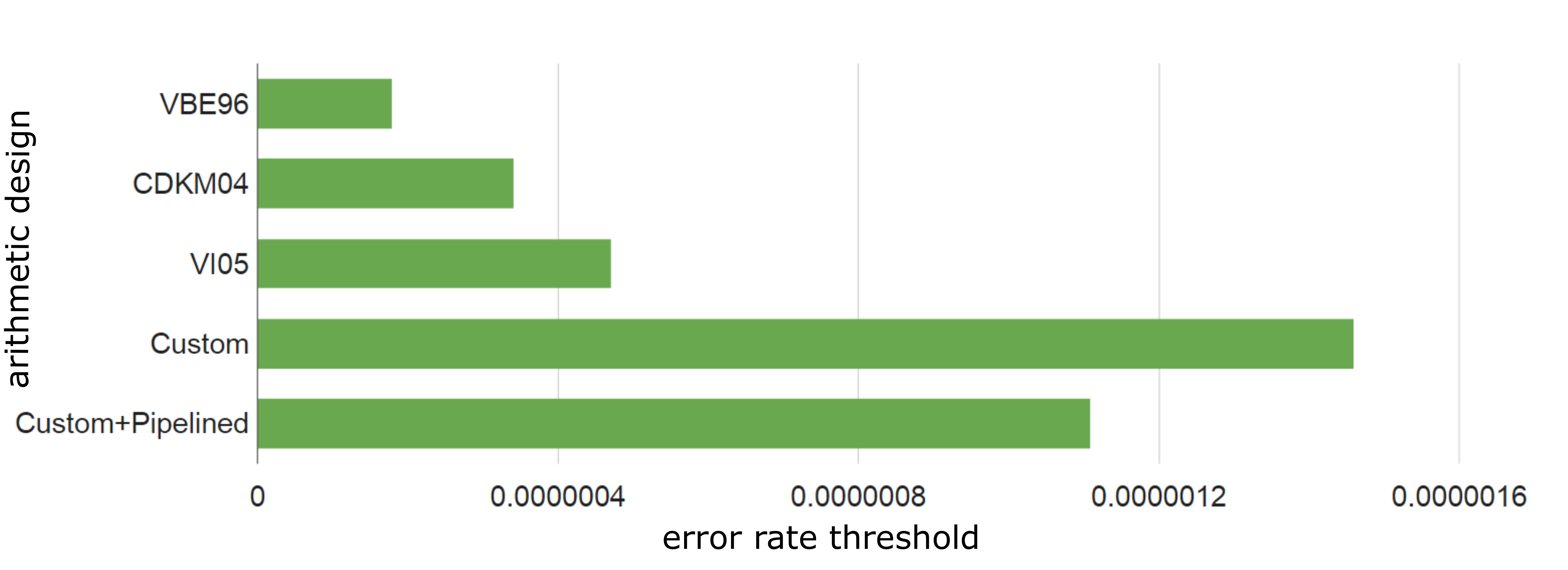}
	\caption{Threshold of the local gate error for the total design. Higher is better.}
	\label{fig:threshold-01}
\end{figure}

We deliberately look to solve the problem at small scale, without employing additional QEC. The attempt is to minimize $KQ$ parameter to ensure that the parameter is significantly less than the error threshold in the emerging technologies.
In this situation, the error threshold for local gates must be considered as an important criterion. We evaluate the error threshold based on the estimation of the product $KQ$. We also employed encoding circuit of the [[5,1,3]] quantum error correction code described in~\cite{Laf96} for applying encoding scheme \scalebox{1.8}{$\varepsilon$}. The result is shown in Fig.~\ref{fig:threshold-01}. 

\begin{table*}
	\centering
		\begin{tabular} {| c || c || c || c |}
		\hline 
		  Phase & Quantum  & Module Circuit & Number of qubits \\ 
			      & Module   & Depth          & (per node)  \\ \hline \hline
			Sharing	&  	 Encoder &  59 & 135\\ 
			Verification & $CX^{b_{i}}$ & 157	& 180 \\ 
			Verification & $QFT$ & 5 & 15 \\ 
		\hline
		\end{tabular}
	\caption{Quantum computational cost of basic quantum circuits and their required qubits during execution of the main phases of graded quantum share and verify (GradedQSV). The [[5,1,3]] quantum error correction code in~\cite{Laf96} is employed for encoding scheme (\scalebox{1.8}{$\varepsilon$}).}
	\label{tab:ComputationalCost}
\end{table*}

\subsection{Network Traffic Cost}
\label{sec:NetworkTrafficCost}
To estimate traffic cost, we divide the result into two sections: The first one is related to the quantum communication cost which is based on transmission of Bell pairs between each pair of nodes in the network. We also consider the cost for sending classical messages during the execution of the protocol. The results are shown in Table~\ref{tab:NetworkCost}. 
 
\begin{table*}
	\centering
		\resizebox{\textwidth}{!} {
		\begin{tabular} {| c || c || c || c || c || c || c |}
		\hline 
		  Main & Phase & End-Node & Source & Destination &	Quantum Cost & Classical Cost\\ 
			Algorithm     &       & Comm Type& Node   &   Node 		 & (Bell Pairs)   & (Bits)\\ \hline \hline
			GradedQSV  		& Sharing	&  	 Unicast &  Dealer & All Nodes & 108						& 0 \\ 
			GradedQSV 		& Sharing & 		 Unicast & Node(i) & Node (j) &  540						& 0\\ 
			GradedQSV 		& Verification & Gradecast & All Nodes & All Nodes &  0						& 14160\\ 
			GradedQSV 		& Verification & Gradecast & All Nodes & All Nodes &  0						& 4720\\ 
			QOCC 			& Measurement &  Gradecast & All Nodes  & All Nodes &  0						& 2360\\ 
		\hline
			total 	& - & - & - & - & 648						& 21240\\
		\hline
		\end{tabular}
		}
	\caption{Quantum traffic cost (Bell pairs) and classical traffic cost during the execution of the quantum operations in QBA. Note that only the classical cost related to the result of quantum blocks is shown and the classical communications for updating list of faulty-nodes is neglected in the table.}
	\label{tab:NetworkCost}
\end{table*}







\section{Conclusion}
\label{sec:Conclusion}
In this paper, we described an optimized quantum architecture for end-nodes in quantum aided Byzantine agreement protocol. The node architecture in the protocol is not as complex as that required for e.g. factoring a large number but at minimum of $165$ qubits per node, the experimental demands are substantial compared to current capabilities. In addition, the number of required Bell pairs is in the order of the number required for quantum key distribution. 

During the design process we found that modular classical-quantum multiplication and fully modular quantum addition are the most critical parts of the quantum computation in the QBA protocol. This requires a novel circuit design that is fully depends on the minimum prime number $P$ satisfy the inequality ($N < P$).  The other requirement is related to long lived qubits to maintain the shared and encoded qubits. 
  
\section*{Acknowledgment} 

This work is supported by JSPS KAKENHI Kiban B 16H02812 and JSPS KAKENHI Grant Number 25280034. The
authors would like to thank Shota Nagayama, Shigeru Yamashita, Shigeya
Suzuki, Takaaki Matsuo and Takahiko Satoh for valuable technical
conversations.

\end{document}